# Electronic Thermal Conductivity and the Wiedemann-Franz Law for Unconventional Superconductors

M. J. Graf, S-K. Yip, and J. A. Sauls

*Department of Physics and Astronomy, Northwestern University, Evanston, Illinois 60208*

D. Rainer

*Physikalisches Institut, Universität Bayreuth, D-95440 Bayreuth, Germany*

(version: 7 September 1995)

We use the quasiclassical theory of superconductivity to calculate the electronic contribution to the thermal conductivity. The theory is formulated for low temperatures when heat transport is limited by electron scattering from random defects and for superconductors with nodes in the order parameter. We show that certain eigenvalues of the thermal conductivity tensor are universal at low temperature, $k_B T \ll \gamma$, where $\gamma$ is the bandwidth of impurity bound states in the superconducting phase. The components of the electrical and thermal conductivity also obey a Wiedemann-Franz law with the Lorenz ratio, $L(T) = \kappa/\sigma T$, given by the Sommerfeld value of $L_S = (\pi^2/3)(k_B/e)^2$ for $k_B T \ll \gamma$. For intermediate temperatures the Lorenz ratio deviates significantly from $L_S$, and is strongly dependent on the scattering cross section, and qualitatively different for resonant vs. nonresonant scattering. We include comparisons with other theoretical calculations and the thermal conductivity data for the high $T_c$ cuprate and heavy fermion superconductors.

PACS number(s): 74.25.Fy, 74.70.Tx, 72.15.Eb

## I. INTRODUCTION

In a normal metal at sufficiently low temperatures the electrical and thermal conductivities are determined by the scattering of electrons by lattice defects. The electrical conductivity approaches a constant, while the heat conductivity, $\kappa(T) \sim T$, is related to the electrical conductivity by Sommerfeld's result for the Lorenz ratio, $\kappa/(\sigma T) \to L_S = (\pi^2/3)(k_B/e)^2$. In fact the Wiedemann-Franz (WF) law is frequently used to estimate the phonon contribution to $\kappa$ by subtracting off the expected electronic contribution, $\sigma L_S T$, from the measured heat conductivity.[1]

Superconductivity has dramatic effects on the electrical and thermal conductivities in conventional (s-wave) superconductors.[2-8] In particular, the WF law is violated by the formation of a coherent ground state and the opening of a gap in the excitation spectrum everywhere on the Fermi surface. In this article we investigate the behavior of the heat current for superconductors with an order parameter of reduced symmetry for which there are gapless excitations even for the pure superconductor. Such superconducting states have been argued to exist both in the cuprates[9] and the heavy fermion systems.[10] In particular, it is widely believed that the order parameter for the heavy fermion superconductor UPt$_3$ vanishes on a line in the basal plane, $p_{fz} = 0$ [e.g., see Ref. 11]. Similarly, one of the leading candidates for the cuprates is the B$_{1g}$ or $d_{x^2-y^2}$ state, a singlet order parameter with lines of zeros at the Fermi surface positions, $p_{fx} = \pm p_{fy}$ [for a review see Ref. 9].

For a *clean* superconductor with an order parameter that vanishes along a line on the Fermi surface the density of states is linear in the excitation energy, $N(\epsilon) \sim N_f \epsilon/\Delta_0$ for $\epsilon < \Delta_0$.[12,13] However, this spectrum is altered by a random distribution of impurities.[14,15] A new energy scale, $\gamma$, develops below which the density of states is approximately constant and non-zero at zero energy. The energy scale $\gamma$ is interpreted as the bandwidth of quasiparticle states bound to impurities.[16-19] These impurity bound states develop below the superconducting transition and are coherent superpositions of particle and hole excitations. Such states are formed by the constructive interference of particle- and hole-like excitations that undergo Andreev scattering from the variations of the order parameter that occur as a result of unconventional pairing and potential scattering by the impurity. For an order parameter with a line of nodes, the bandwidth, $\gamma$, and the density of Andreev bound states at zero energy, $N(0)$, are finite for any finite concentration of impurities, $n_{imp} \neq 0$.[20] The energy scale, $\gamma$, and the density of states, $N(0)$, depend on both the impurity concentration, $n_{imp}$, and the scattering phase shift, $\delta_0$. Thus, $\gamma$ provides a crossover energy scale, below which the transport properties of an unconventional superconductor are dominated by the Andreev bound states. For excitation energies above $\gamma$ the transport properties are determined primarily by the scattering of continuum excitations.

The electrical conductivity for a superconductor with an order parameter that vanishes along a line of nodes was shown theoretically by Lee[21] to have a *universal* limiting value, $\sigma_0 = \lim_{\omega \to 0} \sigma(\omega, T = 0) \simeq e^2 N_f v_f^2 \tau_\Delta$, where $\tau_\Delta \simeq \hbar/(\pi \Delta_0)$ is a universal transport time that is independent of either the concentration or the scattering phase shift. This result was obtained for a two-dimensional (2D) $d_{x^2-y^2}$ order parameter of the form $\Delta(\vec{p}_f) = \Delta_0 \cos(2\varphi)$ and an isotropic 2D Fermi surface, $\vec{p}_f = p_f(\cos\varphi, \sin\varphi)$. Calculations show that the universal value of the conductivity is due to the phase space for "optical" transitions within the band of impurity-induced Andreev bound states.[22,23] The phase space requirements for a universal limit are determined by the variation of the order parameter near the nodes; $\Delta(\vartheta) \sim \vartheta - \vartheta_{node}$ for a line node in 3D or a point node in 2D, and $\Delta(\vartheta) \sim (\vartheta - \vartheta_{node})^2$ for a point node in 3D. The crossover to the universality limit occurs for $k_B T \lesssim \gamma$ and $\hbar\omega \lesssim \gamma$. Thus, the universal limit is most easily realized in the strong scattering (unitarity) limit. Since there is considerable evidence that some of the heavy fermion superconductors have an order parameter with a line of zeroes, the experimental confirmation of the universal result for the conductivity would provide an important test of the argument (based on the Kondo lattice model) that impurity scattering in heavy fermion metals is in the unitarity limit.[13,24] Unitarity scattering by impurities has also been invoked in the high $T_c$ superconductors in order to reconcile the often observed $T^2$ dependence of the penetration depth at low temperatures with a $d$-wave order parameter.[25,26]

Because electromagnetic fields penetrate only a distance of the order of the London length into the superconductor, surface effects can complicate the determination of the bulk conductivity. On the other hand, the heat current is unscreened and provides more direct access to the bulk excitation spectrum. In this article, we investigate the low temperature behavior of the thermal conductivity tensor for unconventional superconductors with line and point nodes in the order parameter. One of the issues we address is whether or not the universal behavior of the electrical conductivity extends to the electronic heat conductivity at low temperature. We show that the components of the electronic thermal conductivity tensor $\overset{\leftrightarrow}{\kappa}$ corresponding to quasiparticles in the vicinity of the line nodes are determined by the same scattering rate as the electrical conductivity and are universal in the limit $T \to 0$. Furthermore, the WF law is obeyed for the ratio of the universal electrical and thermal conductivities in the limit $k_B T \ll \gamma$. However, a significant temperature dependence of the Lorenz ratio occurs over the temperature range, $T < T_c$, even with purely elastic scattering. The universal values for both the electrical and thermal conductivity result from the cancelation between two factors: (i) the density of Andreev bound states, which is proportional to $\gamma$, and (ii) the reduction of phase space for scattering of gapless excitations, which is proportional to $\gamma^{-1}$, leading to an estimate for the thermal conductivity, $\kappa \sim N_f(\gamma/\Delta_0)k_B^2 T v_f^2 (\hbar/\gamma) \sim N_f v_f^2 k_B^2 T(\hbar/\Delta_0)$, which is independent of the defect density or scattering phase shift (again for $n_{imp}^{-1} > \xi_0^3$). Perhaps the most surprising result is that the ratio of the universal values for the thermal and electrical conductivity gives the Sommerfeld value for the Lorenz ratio, $\kappa/\sigma T \simeq L_S = \frac{\pi^2}{3}(k_B/e)^2$. Thus, the differences in the coherence factors that determine the conductivity tensors, $\overset{\leftrightarrow}{\kappa}$ and $\overset{\leftrightarrow}{\sigma}$, do *not* affect the Lorenz ratio $L(T) = \kappa/\sigma T$ for $k_B T \ll \gamma$ and $\hbar\omega \ll \gamma$.

For temperatures above the crossover energy, $k_B T \gtrsim \gamma$, the Lorenz ratio, $L(T)$, deviates significantly from the Sommerfeld value. Furthermore, we find that the temperature dependence of the Lorenz ratio is very sensitive to the scattering phase shift. For nearly resonant scattering [$\delta_0 \approx \pi/2$] $L(T)$ is larger than $L_S$, except in a narrow region near $T_c$ associated with a coherence peak in the electrical conductivity for very clean superconductors. In the opposite limit of weak scattering the Lorenz ratio is less than $L_S$ except for the (exponentially small) region $k_B T \lesssim \gamma$. Thus, measurements of $L(T)$ might be useful in distinguishing weak and strong impurity scattering.

The rest of this paper is organized as follows. In Sec. II we derive an equation for the thermal conductivity of a Fermi liquid; the formulation includes unconventional pairing and the effects of scattering by a random distribution of defects. In Sec. III we evaluate the thermal conductivity tensor in the limit $T \to 0$ for several models of the order parameter: (i) the even-parity $B_{1g}$ state in tetragonal symmetry, i.e., the $d_{x^2-y^2}$ state with $\Delta(\vec{p}_f) = \Delta_0(p_{fx}^2 - p_{fy}^2)$; and in hexagonal systems, (ii) the even-parity $E_{1g}$ ("hybrid-I")[27] state with $\Delta(\vec{p}_f) = 2\Delta_0 p_{fz}(p_{fx} + ip_{fy})$, (iii) the odd-parity $A_{1u}$ ("polar") state with $\vec{\Delta}(\vec{p}_f) = \Delta_0 \hat{z} p_{fz}$, and (iv) the odd-parity $E_{2u}$ ("hybrid-II") state $\vec{\Delta}(\vec{p}_f) = \frac{3\sqrt{3}}{2}\Delta_0 \hat{z} p_{fz}(p_{fx} + ip_{fy})^2$. The $d_{x^2-y^2}$ state has been discussed extensively as a model for the high $T_c$ cuprates, and the latter three order parameters have been discussed as models for the low-temperature superconducting phase of UPt$_3$. For the odd-parity states (cases iii and iv) $\hat{z}$ specifies the quantization direction for the spins; the pairs are in the triplet spin state, $|\downarrow\rangle + |\uparrow\rangle$, relative to the $\hat{z}$ direction. The formulation and many of the results are applicable to more general forms of anisotropic pairing with zeroes in the order parameter. A few other models are also discussed. In Sec. IV we discuss the leading order finite temperature corrections to the thermal and electrical conductivities. For an order parameter with a line of nodes, or a point node in which the gap opens quadratically, the thermal conductivity tensor has components that are universal in the limit $T \to 0$, and of the form $\kappa = L_S \sigma T(1 + \mathcal{O}[T^2/\gamma^2])$, exhibiting both the WF law for $T \to 0$ and the deviations that develop for $k_B T \sim \gamma \ll \Delta_0$. In Sec. V we present numerical results for the thermal conductivity and the Lorenz ratio over the full temperature range below $T_c$, and compare in Sec. VI the results with low-temperature measurements of the thermal conductivity for several cuprate and heavy fermion superconductors.

## II. QUASICLASSICAL TRANSPORT COEFFICIENTS

The microscopic theory of superconductivity was developed by Bardeen, Cooper & Schrieffer[30] at about the same time that Landau published the microscopic basis of his transport theory of normal Fermi liquids.[31] These two theories were combined into what was called by Larkin & Ovchinnikov[32] the *quasiclassical theory* of superconducting Fermi liquids. The quasiclassical theory of su-perconductivity is a complete theory of the low-energy properties of fermions in the superconducting state. It was developed by Eilenberger,[33] Larkin & Ovchinnikov,[32] and Eliashberg,[34] and it covers essentially all thermodynamic, electrodynamic, transport and collective properties of superconductors. In this section we start from the basic equations of the quasiclassical theory of superconductivity, and derive the electrical conductivity and the electronic contribution to the heat conductivity of anisotropic superconductors with unconventional pairing and scattering by random defects.

The central physical objects of the quasiclassical theory are the quasiclassical propagators, which obey quasiclassical transport-like equations. We give a brief interpretation of their physical meaning, and establish our notation. We use Keldysh's formulation of nonequilibrium Green's function theory,[35] and introduce three types of propagators: *advanced (A), retarded (R),* and *Keldysh (K)*. These three quasiclassical propagators are $4\times 4$-matrices whose components describe the quantum-mechanical internal degrees of freedom of electrons and holes: the spin and particle-hole degrees of freedom. Particle and hole excitations are incoherent in the normal state, whereas the superconducting state is characterized by quantum coherence between particles and holes, which is the origin of persistent currents and other nonclassical superconducting effects. The quasiclassical propagators describe the quantum statistical state of the internal degrees of freedom. Nonvanishing off-diagonal elements in the particle-hole index indicate superconductivity. A standard notation for the matrix structure of the propagators (and the self-energies) is[36]

$$\hat{g}^X = \begin{pmatrix} g^X + \vec{g}^X\cdot\vec{\sigma} & \left(f^X + \vec{f}^X\cdot\vec{\sigma}\right)i\sigma_y \\ i\sigma_y\left(\underline{f}^X + \underline{\vec{f}}^X\cdot\vec{\sigma}\right) & \underline{g}^X - \sigma_y\underline{\vec{g}}^X\cdot\vec{\sigma}\sigma_y \end{pmatrix} \quad (1)$$

with $X \in \{R,A,K\}$. The 16 matrix elements of $\hat{g}^X$ are written in terms of four spin scalars ($g^X$, $\underline{g}^X$, $f^X$, $\underline{f}^X$) and four spin vectors ($\vec{g}^X$, $\underline{\vec{g}}^X$, $\vec{f}^X$, $\underline{\vec{f}}^X$). All matrix elements are functions of the Fermi momentum $\vec{p}_f$, the position $\vec{R}$, the excitation energy $\epsilon$, measured from the chemical potential, and the time $t$. The diagonal spin scalars $g^X$, $\underline{g}^X$ contain the spectral and statistical information for spin-independent quantities like the heat current density, while the spin vectors, $\vec{g}^X$, carry the information on the spin magnetization, spin currents, etc. The off-diagonal terms $f^X$ and $\vec{f}^X$ characterize the superconducting state; a finite value of $f^K$ indicates singlet pairing, a nonvanishing $\vec{f}^K$ implies triplet pairing.

The redundant information provided in the definition of the quasiclassical propagators in Eq. (1) can be eliminated with the very general symmetries[36]

$$g^A(\vec{p}_f;\epsilon) = g^R(\vec{p}_f;\epsilon)^*\,,\quad g^R(\vec{p}_f;\epsilon) = \underline{g}^A(-\vec{p}_f;-\epsilon)\,, \quad (2)$$

$$\vec{g}^A(\vec{p}_f;\epsilon) = \vec{g}^R(\vec{p}_f;\epsilon)^*\,,\quad \vec{g}^R(\vec{p}_f;\epsilon) = \underline{\vec{g}}^A(-\vec{p}_f;-\epsilon)\,, \quad (3)$$

$$\underline{f}^A(\vec{p}_f;\epsilon) = f^R(\vec{p}_f;\epsilon)^*\,,\quad f^R(\vec{p}_f;\epsilon) = \underline{f}^A(-\vec{p}_f;-\epsilon)\,, \quad (4)$$

$$\underline{\vec{f}}^A(\vec{p}_f;\epsilon) = \vec{f}^R(\vec{p}_f;\epsilon)^*\,,\quad \vec{f}^R(\vec{p}_f;\epsilon) = -\underline{\vec{f}}^A(-\vec{p}_f;-\epsilon)\,. \quad (5)$$

The electrical current density is obtained from the scalar part of the Keldysh propagator, the Fermi velocity, $\vec{v}_f$, and the density of states per spin, $N_f$, at the Fermi level,

$$\vec{j}_e(\vec{R},t) = 2N_f\int d\vec{p}_f\int\frac{d\epsilon}{4\pi i}\left[e\,\vec{v}_f(\vec{p}_f)\right]g^K(\vec{p}_f,\vec{R};\epsilon,t)\,, \quad (6)$$

where $\int d\vec{p}_f(\ldots)$ stands for a normalized integral over the Fermi surface. Similarly, the heat current density has the form,

$$\vec{j}_\varepsilon(\vec{R},t) = 2N_f\int d\vec{p}_f\int\frac{d\epsilon}{4\pi i}\left[\epsilon\,\vec{v}_f(\vec{p}_f)\right]g^K(\vec{p}_f,\vec{R};\epsilon,t)\,. \quad (7)$$

For weak disturbances from equilibrium the current response is linear in the applied field. In this paper we are interested in the low-frequency, dissipative part of the electrical current response, defined by the conductivity tensor,

$$\delta\vec{j}_e = \overleftrightarrow{\sigma}\cdot\vec{E}_\omega\,, \quad (8)$$

where $\overleftrightarrow{\sigma} = \lim_{\omega\to 0}\mathrm{Re}\,\overleftrightarrow{\sigma}(\omega,T)$,[37] and the thermal conductivity tensor, defined by the linear response to a small temperature gradient

$$\delta\vec{j}_\varepsilon = -\overleftrightarrow{\kappa}\cdot\vec{\nabla}T\,. \quad (9)$$

In the rest of this section we develop the linear response equations for $\overleftrightarrow{\kappa}$ and $\overleftrightarrow{\sigma}$ from the quasiclassical theory. The analysis and notation closely follows that for the current response to an EM field given in Refs. 38 and 22. The advanced, retarded, and Keldysh propagators are calculated from quasiclassical transport equations,

$$\left[\epsilon\hat{\tau}_3 - \hat{\sigma}_{ext} - \hat{\sigma}^{R,A}\,,\,\hat{g}^{R,A}\right]_\circ + i\vec{v}_f\cdot\vec{\nabla}\hat{g}^{R,A} = 0\,, \quad (10)$$

and

$$\left(\epsilon\hat{\tau}_3 - \hat{\sigma}_{ext} - \hat{\sigma}^R\right)\circ\hat{g}^K - \hat{g}^K\circ\left(\epsilon\hat{\tau}_3 - \hat{\sigma}_{ext} - \hat{\sigma}^A\right)$$
$$-\hat{\sigma}^K\circ\hat{g}^A + \hat{g}^R\circ\hat{\sigma}^K + i\vec{v}_f\cdot\vec{\nabla}\hat{g}^K = 0\,, \quad (11)$$

where all propagators and self-energies depend on the Fermi momentum $\vec{p}_f$, the position $\vec{R}$, the excitation energy $\epsilon$, and the time $t$. We use the units $\hbar \equiv k_B \equiv 1$, unless explicitly stated. The $\circ$-product stands for the following operation in the energy-time variables

$$\left(\hat{a}\circ\hat{b}\right)(\vec{p}_f,\vec{R};\epsilon,t) =$$
$$e^{\frac{i}{2}\left(\partial^a_\epsilon\partial^b_t - \partial^a_t\partial^b_\epsilon\right)}\hat{a}(\vec{p}_f,\vec{R};\epsilon,t)\hat{b}(\vec{p}_f,\vec{R};\epsilon,t)\,, \quad (12)$$

and the commutator $[\hat{a},\hat{b}]_\circ$ is defined by $\hat{a}\circ\hat{b} - \hat{b}\circ\hat{a}$. The transport equations are supplemented by the normalization conditions,[33,32]

$$\hat{g}^{R,A}\circ\hat{g}^{R,A} = -\pi^2\hat{1}, \quad (13)$$

$$\hat{g}^R\circ\hat{g}^K + \hat{g}^K\circ\hat{g}^A = 0\,. \quad (14)$$

The quasiclassical transport equations (10)-(11) together with the normalization conditions (13)-(14) and the equations specifying the self-energies, $\hat{\sigma}^X$, are the fundamental equations of the Fermi liquid theory of superconductivity. They are the generalization of the Boltzmann-Landau transport equation to the superconducting state.

The transformations and approximations used to derive transport equations are based on a systematic expansion to leading order in the small parameters of Fermi liquid theory, e.g., $k_B T_c/E_f$, $\hbar\omega_D/E_f$, $\hbar/\tau E_f$, etc. The accuracy and predictive power of Fermi liquid theory is intimately connected with the smallness of these parameters.

The quasiclassical self-energy terms, $\hat{\sigma}^X$, in the transport equations describe interactions between quasiparticles with phonons, with impurities, and quasiparticles with each other. We consider the low temperature transport properties of superconductors with unconventional pairing under conditions where inelastic scattering by phonons and quasiparticles is negligible compared to scattering from random defects.

The quasiclassical self-energies depend on interaction vertices which are phenomenological parameters of the Fermi liquid theory of superconductivity. We consider the weak-coupling limit[38] in which electronic pairing interactions are described by the vertices $V^s_{\vec{p}_f \vec{p}_f{'}}$ and $V^t_{\vec{p}_f \vec{p}_f{'}}$, for the spin-singlet and spin-triplet interactions, respectively. The mean-field pairing self-energies are given by

$$\Delta^{R,A}(\vec{p}_f, \vec{R}; t) = \int \frac{d\epsilon}{4\pi i} \int d\vec{p}_f{'} V^s_{\vec{p}_f \vec{p}_f{'}} f^K(\vec{p}_f{'}, \vec{R}; \epsilon, t), \quad (15)$$

$$\vec{\Delta}^{R,A}(\vec{p}_f, \vec{R}; t) = \int \frac{d\epsilon}{4\pi i} \int d\vec{p}_f{'} V^t_{\vec{p}_f \vec{p}_f{'}} \vec{f}^K(\vec{p}_f{'}, \vec{R}; \epsilon, t), \quad (16)$$

$$\Delta^K = \vec{\Delta}^K = 0. \quad (17)$$

The effects of a random distribution of impurities are described in Fermi liquid theory by an electron-impurity vertex, $u(\vec{p}_f, \vec{p}_f{'})$, and the impurity concentration $n_{imp}$.[20] The impurity self-energy is proportional to the single impurity $\hat{t}$ matrix,

$$\hat{\sigma}^X_{imp}(\vec{p}_f, \vec{R}; \epsilon, t) = n_{imp} \hat{t}^X(\vec{p}_f, \vec{p}_f, \vec{R}; \epsilon, t), \quad (18)$$

where $\hat{t}^X$ are obtained by solving the integral equations,

$$\hat{t}^{R,A}(\vec{p}_f, \vec{p}_f{'}, \vec{R}; \epsilon, t) = u(\vec{p}_f, \vec{p}_f{'}) \hat{1} + N_f \int d\vec{p}_f{''} u(\vec{p}_f, \vec{p}_f{''})$$
$$\hat{g}^{R,A}(\vec{p}_f{''}, \vec{R}; \epsilon, t) \circ \hat{t}^{R,A}(\vec{p}_f{''}, \vec{p}_f{'}, \vec{R}; \epsilon, t), \quad (19)$$

$$\hat{t}^K(\vec{p}_f, \vec{p}_f{'}, \vec{R}; \epsilon, t) = N_f \int d\vec{p}_f{''} \hat{t}^R(\vec{p}_f, \vec{p}_f{''}, \vec{R}; \epsilon, t)$$
$$\circ \hat{g}^K(\vec{p}_f{''}, \vec{R}; \epsilon, t) \circ \hat{t}^A(\vec{p}_f{''}, \vec{p}_f{'}, \vec{R}; \epsilon, t). \quad (20)$$

The coupling of quasiparticles (with charge $e$) to an electric field is given by

$$\hat{\sigma}_{ext} = -\frac{e}{c} \vec{v}_f \cdot \vec{A} \hat{\tau}_3, \quad (21)$$

where $\vec{A}(\vec{q},\omega)$ is the vector potential describing the transverse field $\vec{E} = \frac{i\omega}{c}\vec{A}$. In order to calculate the conductivity we must solve the transport equations to linear order in the perturbing field. We first linearize the transport equations in the perturbations of $\hat{g}^X$ and $\hat{\sigma}^X$ from their equilibrium values. In the case of the heat transport the perturbation is the temperature gradient $\vec{\nabla}T$. Thus, the linearization of the transport equations is carried out in terms of the deviations from *local* equilibrium specified by a thermal distribution function with a local temperature,

$$\Phi_0(\vec{R}) = [1 - 2f(\epsilon; T(\vec{R}))] = \tanh\left(\frac{\epsilon}{2T(\vec{R})}\right). \quad (22)$$

The local equilibrium Keldysh propagator and self-energy are determined by the retarded and advanced functions and the thermal distribution function,

$$\hat{g}^K_0 = \hat{g}^R_0 \circ \Phi_0 - \Phi_0 \circ \hat{g}^A_0, \quad (23)$$
$$\hat{\sigma}^K_0 = \hat{\sigma}^R_0 \circ \Phi_0 - \Phi_0 \circ \hat{\sigma}^A_0, \quad (24)$$

with retarded and advanced propagators that are given by the solutions of

$$\left[\epsilon\hat{\tau}_3 - \hat{\sigma}^{R,A}_0, \hat{g}^{R,A}_0\right] = 0, \quad (25)$$

$$\left[\hat{g}^{R,A}_0\right]^2 = -\pi^2 \hat{1}, \quad (26)$$

and the self-consistency equations (15)-(20). Note, that the o-product reduces to matrix multiplication for the local equilibrium functions. The self-energy includes the mean-field order parameter, $\hat{\Delta}(\vec{p}_f, \vec{R})$, and the impurity self-energy, $\hat{\sigma}^{R,A}_{imp}(\vec{p}_f, \vec{R}; \epsilon)$, which has both diagonal ($\hat{\Sigma}^{R,A}_{imp}$) and off-diagonal ($\hat{\Delta}^{R,A}_{imp}$) components in particle-hole space.

In this paper we consider only superconducting states which are "unitary", i.e., the equilibrium mean-field order parameter satisfies

$$\hat{\Delta}(\vec{p}_f, \vec{R})^2 = -|\Delta(\vec{p}_f)|^2 \hat{1}, \quad (27)$$

where $|\Delta|^2$ stands for either the spin scalar product $\Delta \underline{\Delta}$ or the spin vector product $\vec{\Delta} \cdot \underline{\vec{\Delta}}$. The unitary condition restricts us to even-parity, spin-singlet pairing or to odd-parity, spin-triplet states without spontaneous spin polarization. The odd-parity states considered in this paper are unitary states that do not break time-reversal symmetry in the spin degrees of freedom. However, time-reversal symmetry may still be broken by the orbital motion of the Cooper pairs, which is the case for the $E_{1g}$ and $E_{2u}$ ground states that we consider.[39]

The local equilibrium solutions to Eqs. (25)-(26) for unitary states in unconventional superconductors [with $\int d\vec{p}_f \hat{\Delta}(\vec{p}_f) = 0$] are[40]

$$\hat{g}^{R,A}_0(\vec{p}_f, \vec{R}; \epsilon) = -\pi \frac{\tilde{\epsilon}^{R,A}\hat{\tau}_3 - \hat{\Delta}}{\sqrt{|\Delta|^2 - (\tilde{\epsilon}^{R,A})^2}}, \quad (28)$$

$$\tilde{\epsilon}^{R,A}(\vec{p}_f, \vec{R}; \epsilon) = \epsilon - \frac{1}{4} \text{Tr}\left[\hat{\tau}_3 \hat{\Sigma}^{R,A}_{imp}(\vec{p}_f, \vec{R}; \epsilon)\right]. \quad (29)$$

These equilibrium functions are inputs to the linearized quasiclassical transport equations. The quasiclassical transport equations and normalization conditions are solved to linear order for the deviation of the propagators from their local equilibrium values, $\delta\hat{g}^X(\vec{p}_f, \vec{R}; \epsilon, t) = \hat{g}^X(\vec{p}_f, \vec{R}; \epsilon, t) - \hat{g}^X_0(\vec{p}_f, \vec{R}; \epsilon)$. The technical steps used to decouple the retarded, advanced and Keldysh functions, and for inverting the linearized transport equations are outlined in the Appendix, and the solution for $\delta\hat{g}^K$ is given in Eq. (A20). In the following we use the general solution for $\delta\hat{g}^K$ and the self-consistency equations for the impurity self-energy and order parameter to obtain formulas for the electrical and thermal conductivities for a superconductor with an unconventional order parameter.

## III. ELECTRICAL AND THERMAL CONDUCTIVITIES

Here we consider a superconductor with anisotropic singlet pairing or unitary triplet pairing, and discuss the electrical and thermal conductivities in the long wavelength limit $q \to 0$ and at $T \to 0$. For simplicity we assume isotropic impurity scattering. In this case the first order corrections to the current response functions from the impurity self-energy, $\delta\hat{\sigma}_{imp}$, and the order parameter, $\delta\hat{\Delta}$, vanish for all listed pairing states, except for the polar state (ii) with current flow along the c-axis.[15,41,42] Self-energy corrections corresponding to the excitation of collective modes of the order parameter, $\delta\hat{\Delta}$, also vanish in the limit $q \to 0$ [cf. Refs. 43 and 44]. The self-energy corrections are the 'vertex corrections' in the language of the Green's function formulation of the Kubo response function.[45] If we can neglect vertex corrections, we obtain expressions for the electrical and thermal conductivities that depend only on the equilibrium propagators and self-energies, and the external perturbations. For spin singlet states the thermal conductivity obtained from Eqs. (7) and (A20) becomes

$$\kappa_{ij}(T) = -\frac{N_f}{4\pi T^2} \int d\epsilon \int d\vec{p}_f \, [v_{f,i} v_{f,j}] \, \epsilon^2 \text{sech}^2\left(\frac{\epsilon}{2T}\right)$$
$$\times \frac{C_+^a(\vec{p}_f;\epsilon)}{\pi^2 C_+^a(\vec{p}_f;\epsilon)^2 + D_-^a(\epsilon)^2} \left[g_0^A(\vec{p}_f;\epsilon) g_0^R(\vec{p}_f;\epsilon)\right.$$
$$\left. - \underline{f}_0^A(\vec{p}_f;\epsilon) f_0^R(\vec{p}_f;\epsilon) + \pi^2 \right]. \quad (30)$$

In case of spin triplet pairing the off-diagonal spin scalar Green's functions in (30) have to be replaced by $\underline{f}_0^A f_0^R \to \underline{\vec{f}}_0^A \cdot \vec{f}_0^R$. Note, that only the anomalous part of the propagator (see Appendix) contributes to the thermal conductivity. The retarded and advanced parts drop out after taking the trace and applying the normalization condition, i.e., $\text{Tr}\, \hat{g}_0^{R,A} \vec{\nabla} \hat{g}_0^{R,A} = 0$. Physically, this means that the deviation of the quasiparticle distribution function due to a thermal gradient contributes to the heat current, whereas changes in the quasiparticle and Cooper pair spectrum do not. Equation (30), combined with the equilibrium propagators, impurity self-energy and order parameter, is the basic result for the electronic contribution to the thermal conductivity tensor. Note, that we have used the short-hand notation: $C_+^{R,A,a}(\vec{p}_f;\epsilon)$, and $D_-^{R,A,a}(\epsilon)$, for the functions in Eq. (30) at $\omega = 0$. It can be shown that Eq. (30) for $\kappa_{ij}$ reduces to the same expression for the thermal conductivity as reported previously by Schmitt-Rink et al.,[24] Hirschfeld et al.,[46,42] and by Fledderjohann & Hirschfeld;[47] except that these authors appear to have dropped the $D_-^a$ term from the impurity self-energy, which, however, vanishes in both Born and unitarity limits.

Similarly, the electrical conductivity obtained from Eqs. (6) and (A20) for a spin singlet state is given by

$$\text{Re}\,\sigma_{ij}(\omega,T) = \frac{e^2 N_f}{2\pi\omega} \int d\epsilon \int d\vec{p}_f \, [v_{f,i} v_{f,j}] \left[\tanh\left(\frac{\epsilon_+}{2T}\right) - \tanh\left(\frac{\epsilon_-}{2T}\right)\right]$$
$$\times \text{Re}\left\{\frac{C_+^R(\vec{p}_f;\epsilon,\omega)}{\pi^2 C_+^R(\vec{p}_f;\epsilon,\omega)^2 + D_-^R(\epsilon,\omega)^2} \left(g_0^R(\vec{p}_f;\epsilon_-) g_0^R(\vec{p}_f;\epsilon_+) + \underline{f}_0^R(\vec{p}_f;\epsilon_-) f_0^R(\vec{p}_f;\epsilon_+) + \pi^2\right)\right.$$
$$\left. - \frac{C_+^a(\vec{p}_f;\epsilon,\omega)}{\pi^2 C_+^a(\vec{p}_f;\epsilon,\omega)^2 + D_-^a(\epsilon,\omega)^2} \left(g_0^A(\vec{p}_f;\epsilon_-) g_0^R(\vec{p}_f;\epsilon_+) + \underline{f}_0^A(\vec{p}_f;\epsilon_-) f_0^R(\vec{p}_f;\epsilon_+) + \pi^2\right)\right\}, \quad (31)$$

where $\epsilon_\pm = \epsilon \pm \omega/2$. For triplet pairing simply replace $\underline{f}_0(\epsilon_-)f_0(\epsilon_+) \to \underline{\vec{f}}_0(\epsilon_-) \cdot \vec{f}_0(\epsilon_+)$. This result was obtained earlier for magnetic scattering in conventional superconductors,[48] and for electron-phonon and impurity scattering in strong coupling superconductors,[38] and for the in-plane conductivity of layered superconductors.[22] The formula for the conductivity reduces to the well-known result of Mattis and Bardeen for dirty, s-wave superconductors,[2] and to the result derived in Ref. 49 for the frequency and temperature dependence of the conductivity of a d-wave superconductor with lines of nodes in the order parameter.

In deriving Eqs. (30) and (31) we have made use of several relations which are consequences of the general symmetries of the propagator (and self-energy) in Eqs. (2)-(5), the parity of the order parameter, and the specific symmetry of the equilibrium Green's function $\underline{g}_0^X = -g_0^X$, which follows directly from Eq. (28). The basic functions defining the self-energy and response functions obey the following symmetries[36,38]

$$\tilde{\epsilon}^A(\epsilon) = \tilde{\epsilon}^R(\epsilon)^* \,, \quad \tilde{\epsilon}^A(\epsilon) = -\tilde{\epsilon}^R(-\epsilon) \,, \quad (32)$$
$$C_+^R(\vec{p}_f;\epsilon,\omega) = C_+^A(\vec{p}_f;\epsilon,\omega)^* \,, \quad (33)$$
$$C_+^a(\vec{p}_f;\epsilon,\omega) = C_+^a(\vec{p}_f;\epsilon,-\omega)^* \,, \quad (34)$$
$$D_-^R(\epsilon,\omega) = D_-^A(\epsilon,\omega)^* \,, \quad (35)$$
$$D_-^a(\epsilon,\omega) = -D_-^a(\epsilon,-\omega)^* \,. \quad (36)$$

## A. Wiedemann-Franz Law for $T \to 0$

In the limit $T \to 0$ and $\omega \to 0$ the occupation factors $[\tanh((\epsilon+\frac{\omega}{2})/2T) - \tanh((\epsilon-\frac{\omega}{2})/2T)]$ and $\text{sech}^2(\epsilon/2T)$ confine the $\epsilon$-integrals in Eqs. (30)-(31) to a small $\epsilon$ region (of order $T$ or $\omega$). Assuming that there exists an energy scale, $\epsilon^* \gg T$, on which the propagators and self-energies vary, we can set $\epsilon = 0$ in the slowly varying parts of the integrands and obtain

$$\text{Re}\,\sigma_{ij}(\omega \to 0, T \to 0) = \lim_{\omega \to 0} \frac{e^2 N_f}{2\pi\omega} \sinh\frac{\omega}{2T} \int d\epsilon\,\text{sech}^2\left(\frac{\epsilon}{2T}\right)$$

$$\times \text{Re}\int d\vec{p}_f\,[v_{f\,i}v_{f\,j}]\left(\frac{C^R_+(\vec{p}_f)}{\pi^2 C^R_+(\vec{p}_f)^2 + D^R_-(0)^2}\left(g_0^R(\vec{p}_f)g_0^R(\vec{p}_f) + \underline{f}_0^R(\vec{p}_f)f_0^R(\vec{p}_f) + \pi^2\right)\right.$$

$$\left. - \frac{C^a_+(\vec{p}_f)}{\pi^2 C^a_+(\vec{p}_f)^2 + D^a_-(0)^2}\left(g_0^A(\vec{p}_f)g_0^R(\vec{p}_f) + \underline{f}_0^A(\vec{p}_f)f_0^R(\vec{p}_f) + \pi^2\right)\right), \quad (37)$$

and

$$\kappa_{ij}(T \to 0) = -\frac{N_f}{4\pi T^2}\int d\epsilon\,\epsilon^2 \text{sech}^2\left(\frac{\epsilon}{2T}\right)$$

$$\times \int d\vec{p}_f\,[v_{f\,i}v_{f\,j}]\frac{C^a_+(\vec{p}_f)}{\pi^2 C^a_+(\vec{p}_f)^2 + D^a_-(0)^2}\left(g_0^A(\vec{p}_f)g_0^R(\vec{p}_f) - \underline{f}_0^A(\vec{p}_f)f_0^R(\vec{p}_f) + \pi^2\right), \quad (38)$$

where the energies and frequencies are fixed to zero in the arguments of the propagators and self-energies. Using the symmetry relations (2)-(5) and (32)-(36) and eliminating the advanced and anomalous functions in Eqs. (37) and (38), we find

$$\text{Re}\,\sigma_{ij}(\omega \to 0, T \to 0) = \frac{e^2 N_f}{2\pi T}\int d\epsilon\,\text{sech}^2\left(\frac{\epsilon}{2T}\right)\int d\vec{p}_f\,[v_{f,i}v_{f,j}]\frac{g_0^R(\vec{p}_f)^2}{\pi^2 C^R_+(\vec{p}_f)}, \quad (39)$$

and

$$\kappa_{ij}(T \to 0) = \frac{N_f}{2\pi T^2}\int d\epsilon\,\epsilon^2\,\text{sech}^2\left(\frac{\epsilon}{2T}\right)\int d\vec{p}_f\,[v_{f,i}v_{f,j}]\frac{g_0^R(\vec{p}_f)^2}{\pi^2 C^R_+(\vec{p}_f)}. \quad (40)$$

We used the normalization condition, $g_0^R(\vec{p}_f)^2 - \underline{f}_0^R(\vec{p}_f)f_0^R(\vec{p}_f) = -\pi^2$, in addition to the symmetry relations, in order to put the momentum integrals in Eqs. (39) and (40) in identical form. It is useful to write our final results in terms of an average Fermi velocity and a tensor, $\tau_{ij}$, which incorporates all of the coherence effects of superconductivity at $T \to 0$ into an *effective* transport scattering time. The energy integrals are standard, so the conductivities for a system with $D$ dimensions reduce to

$$\text{Re}\,\sigma_{ij}(\omega \to 0, T \to 0) = e^2\frac{2}{D}N_f v_f^2\,\tau_{ij}, \quad (41)$$

$$\kappa_{ij}(T \to 0) = \frac{\pi^2}{3}\frac{2}{D}N_f v_f^2\,T\,\tau_{ij}, \quad (42)$$

where $v_f^2 = \int d\vec{p}_f |\vec{v}_f(\vec{p}_f)|^2$, and the effective transport time is defined by the tensor

$$\tau_{ij} = -\frac{D}{2v_f^2}\int d\vec{p}_f\,\frac{[v_{f,i}(\vec{p}_f)v_{f,j}(\vec{p}_f)]\,\tilde{\epsilon}^R(0)^2}{[\Delta^R(\vec{p}_f)^2 - \tilde{\epsilon}^R(0)^2]^{\frac{3}{2}}}. \quad (43)$$

For an isotropic normal metal one has $\tilde{\epsilon}^R(0) = i/2\tau_0$, where $\tau_0$ is the quasiparticle lifetime due to impurity scattering in the normal state. The transport lifetime in the normal state reduces to $\tau_0$ for isotropic impurity scattering, i.e., $\tau_{ij} = \tau_0 \delta_{ij}$. Note, that Eq. (43) is applicable to the normal state because the key assumption in deriving Eqs. (39)-(40) was that $T$ is small compared with $\epsilon^*$, where $\epsilon^*$ is the energy on which the propagators and self-energies vary. Thus for the normal state $\epsilon^* \sim E_f$, while for the superconducting state $\epsilon^* \sim \gamma$, where $\gamma$ is the impurity bandwidth. In some respect the impurity band states form a new low temperature metallic state deep in the superconducting phase. This analogy is strengthened later in the paper when we calculate the temperature corrections to the transport coefficients using a Sommerfeld expansion. However, the "metallic" band of impurity states has other properties that differ significantly from those of conventional metals. The special features of the impurity induced metallic band reflect the reduced dimensionality for the phase space of scattering and the energy dependence of the particle-hole coherence factors, which define the impurity induced band. These two features lead to (i) universality for $T \to 0$ transport coefficients for excitation gaps with line nodes or quadratic point nodes, and (ii) to the temperature dependence of the Lorenz ratio for elastic scattering at $0 < T < T_c$.

However for $T \to 0$, we note that one obtains from Eqs. (41) and (42) the Wiedemann-Franz law with Sommerfeld's result for the Lorenz number of an unconventional superconductor at very low temperatures, $\kappa_{ij} = L_S\,\sigma_{ij}\,T$, with

$$L_S = \frac{\pi^2}{3}\left(\frac{k_B}{e}\right)^2. \qquad (44)$$

We emphasize that Eqs. (41), (42) hold for gapless superconductors in which the leading contribution to the transport current is that from quasiparticle excitations with energies $T < \gamma \ll T_c$. For superconductors with a gap the number of quasiparticle excitations at low temperature is activated, $\propto \exp(-\Delta_0/T)$, and the transport coefficients in the limit $T \ll T_c$ are not described by Eqs. (41), (42), and furthermore do not obey the Wiedemann-Franz law.

### B. Universal Limits for $d$-wave pairing

Lee has shown that for $T \to 0$ the electrical conductivity of a $d$-wave superconductor is universal,[21]

$$\sigma_\parallel(\omega \to 0, T \to 0) \simeq e^2 N_f v_f^2 \tau_\Delta, \qquad (45)$$

with the universal transport scattering time

$$\tau_\Delta \simeq \frac{1}{\pi \Delta_0}, \qquad (46)$$

which is independent of the scattering rate of quasiparticle excitations.

Our quasiclassical calculation reproduces Lee's result for isotropic systems, $\vec{p}_f = p_f(\cos\varphi, \sin\varphi)$, and the standard model for $d$-wave pairing, i.e., $\Delta(\vec{p}_f) = \Delta_0 \cos(2\varphi)$. We obtain the universal result in the limit, $|\tilde{\epsilon}^R(0)| \ll \Delta_0$, where $\Delta_0$ is the maximum gap. The transport lifetime tensor from Eq. (43) for the standard $d$-wave model reduces to $\tau_{ij} = \tau\,\delta_{ij}$,

$$\tau = \int \frac{d\varphi}{2\pi} \frac{\cos^2(\varphi)\,\gamma^2}{[\Delta(\varphi)^2 + \gamma^2]^{\frac{3}{2}}}, \qquad (47)$$

where $\gamma$ is the width of the impurity band,

$$\gamma = \Gamma_u \frac{\langle \gamma(\Delta^2+\gamma^2)^{-1/2}\rangle}{\cot^2\delta_0 + \langle\gamma(\Delta^2+\gamma^2)^{-1/2}\rangle^2}, \qquad (48)$$

and $<\ldots> = \int_0^{2\pi}\frac{d\varphi}{2\pi}(\ldots)$, and $\Gamma_u = n_{imp}/(\pi N_f)$. For a given impurity concentration this bandwidth is largest in the limit of unitarity scattering, e.g., for $d_{x^2-y^2}$ pairing, $\gamma \sim \sqrt{\pi\Delta_0\Gamma_u/2}$, while in the Born limit, $\gamma \sim 4\Delta_0 \exp(-\pi\Delta_0/2\Gamma_u)$. In either case we have $\gamma \ll \Delta_0$, otherwise pair-breaking by impurity scattering effectively suppresses the superconducting transition. For $\gamma \ll \Delta_0$, Eq. (47) reduces to $\tau \simeq 1/(\pi\Delta_0)$, and we obtain Lee's universal result in Eq. (45).

Since, according to Eq. (43), the same transport lifetime determines the electronic contribution to the thermal conductivity, it too becomes universal in the limit $T \to 0$, i.e., independent of the impurity scattering rates. The universal, low-temperature limits for the electrical and thermal conductivities of a standard $d$-wave superconductor with $\Delta \propto \cos(2\varphi)$ and scattering due to isotropic impurities also obey the WF law. Although scattering by impurities is expected to be the dominant scattering mechanism at low temperatures, it is worth noting that these universal results also hold for electron-electron scattering or electron-phonon scattering, *as long as vertex corrections can be neglected*.

## IV. LOW TEMPERATURE LIMITS FOR SEVERAL UNCONVENTIONAL PAIRING STATES

### A. Zero temperature limit

We evaluate the eigenvalues of the $\overleftrightarrow{\kappa}$'s in the $T \to 0$ limit for the specific pairing states listed in Table I for uniaxial superconductors. For heat flow in the basal plane the two eigenvalues of $\kappa_{ij}$ are identical, and we drop the subscripts for these directions and write $\kappa_\parallel = \kappa_{aa} = \kappa_{bb}$, and $\kappa_\perp = \kappa_{cc}$ for the $c$-axis transport coefficient.

First, consider the $d_{x^2-y^2}$ pairing state. For a cylindrical Fermi surface and the heat flow in the $ab$-plane, the angular average in Eq. (47) for the transport time reduces to $\frac{1}{2}\cdot 4 \cdot \int_{node}\frac{d\varphi}{2\pi}\gamma^2(\Delta^2+\gamma^2)^{-3/2}$. There are four line nodes and the integral is reduced to one quadrant containing one node. For $\gamma \ll \Delta_0$, the integral is dominated by the region near the node, thus, we can approximate $\Delta(\varphi) \simeq \mu\Delta_0\varphi'$, where $\varphi' = \pi/4 - \varphi$ and the parameter $\mu$ measures the slope of the gap at the node, i.e., $\mu = (1/\Delta_0)|d\Delta(\varphi)/d\varphi|_{\varphi_{node}}$. This slope parameter also determines the low-energy density of states [*cf.* Ref. 50]. The result for the low-temperature limit of the thermal conductivity is universal but depends on the slope parameter $\mu$,

$$\kappa_\parallel(T) \simeq \frac{\pi^2}{3} N_f v_f^2 T \frac{2}{\pi\mu\Delta_0}. \qquad (49)$$

The dependence of the universal value of the thermal conductivity on the slope of the excitation gap reflects the importance of the low-energy continuum states with $\epsilon \ll \Delta_0$ in the formation of the zero-energy impurity bound states. Note, that this result is valid for s-wave scattering and $\gamma \ll \Delta_0$ with corrections typically of the order $(\gamma/\Delta_0)^2$. A similar calculation for the in-plane conductivity gives $\mathrm{Re}\,\sigma_\parallel(T,\omega \to 0) \simeq 2e^2 N_f v_f^2/(\pi\mu\Delta_0)$, in agreement with Lee for the standard $d$-wave model with $\mu = 2$.

For the three pairing states listed in Table I containing a line node in the basal plane (polar, hybrid-I, hybrid-II), the relevant angular average for heat flow in the basal plane is $\langle\gamma^2\hat{v}_{f_x}^2(\Delta^2+\gamma^2)^{-3/2}\rangle$ with $\vec{v}_f = v_f\,\hat{v}_f$, which becomes $\frac{1}{4}\cdot\int_0^\pi d\vartheta\,\sin^3\vartheta\,\gamma^2(\Delta^2+\gamma^2)^{-3/2}$. The integral is dominated by the contribution near the line node, i.e., $\vartheta \approx \pi/2$. For $\gamma \ll \Delta_0$, we linearize the gap in the neighborhood of the line node, $\Delta(\vartheta) \simeq \mu\Delta_0(\pi/2 - \vartheta)$, and the integral reduces to $\frac{1}{4}\cdot\int_{node}d\vartheta'\,\gamma^2[(\mu\Delta_0\vartheta')^2+\gamma^2]^{-3/2}$, again leading to a universal result for $\kappa_\parallel$,

TABLE I. Symmetry classes, order parameters, and the asymptotic values of the thermal conductivity tensor. Note, that we have neglected the vertex corrections only for $\kappa_\perp$ of the polar state.

| Pairing State | Symmetry Class (Group) | $\Delta(\vec{p}_f)$ | Nodes | $\dfrac{\kappa_\parallel(T)}{T}\left(\dfrac{\pi^2}{3}N_f v_{f\parallel}^2\right)^{-1}$ | $\dfrac{\kappa_\perp(T)}{T}\left(\dfrac{\pi^2}{3}N_f v_{f\perp}^2\right)^{-1}$ |
|---|---|---|---|---|---|
| $d_{x^2-y^2}$ | $B_{1g}$ ($D_{4h}$) | $(p_{fx}^2 - p_{fy}^2)$ | 4 linear line nodes | $\dfrac{2}{\pi\mu\Delta_0}$ | — |
| polar | $A_{1u}$ ($D_{6h}$) | $\hat{z}\, p_{fz}$ | 1 linear line node | $\dfrac{1}{2\mu\Delta_0}$ | $\dfrac{\sim 1}{\mu\Delta_0}\left(\dfrac{\gamma}{\mu\Delta_0}\right)^2 \ln\dfrac{\mu\Delta_0}{\gamma}$ |
| hybrid-I | $E_{1g}$ ($D_{6h}$) | $p_{fz}(p_{fx}+ip_{fy})$ | 2 linear point nodes + 1 linear line node | $\dfrac{1}{2\mu\Delta_0}$ | $\dfrac{\gamma}{\mu_1^2\Delta_0^2}$ |
| hybrid-II | $E_{2u}$ ($D_{6h}$) | $\hat{z}\, p_{fz}(p_{fx}+ip_{fy})^2$ | 2 quadratic point nodes + 1 linear line node | $\dfrac{1}{2\mu\Delta_0}$ | $\dfrac{1}{2\mu_2\Delta_0}$ |
| hybrid-III-A | $E_{1g}$ ($D_{6h}$) | $p_{fz}(p_{fx}+ip_{fy})(p_{fx}^2+p_{fy}^2)$ | 2 cubic point nodes + 1 linear line node | $\dfrac{1}{2\mu\Delta_0}$ | $\dfrac{0.47}{\mu_3\Delta_0}\left(\dfrac{\mu_3\Delta_0}{\gamma}\right)^{\frac{1}{3}}$ |
| hybrid-III-B | $B_{2g}+iB_{1g}$ ($D_{6h}$) | $p_{fz}(p_{fx}+ip_{fy})^3$ | 2 cubic point nodes + 1 linear line node | $\dfrac{1}{2\mu\Delta_0}$ | $\dfrac{0.47}{\mu_3\Delta_0}\left(\dfrac{\mu_3\Delta_0}{\gamma}\right)^{\frac{1}{3}}$ |
| — | $B_{1u}$ ($D_{6h}$) | $\hat{z}\,\mathrm{Im}(p_{fx}+ip_{fy})^3$ | 2 cubic point nodes + 3 linear line nodes | $\dfrac{3}{2\mu\Delta_0}$ | $\dfrac{\sim 10}{\mu_3\Delta_0}\left(\dfrac{\mu_3\Delta_0}{\gamma}\right)^{\frac{1}{3}}$ |

$$\kappa_\parallel(T) \simeq \frac{\pi^2}{3}N_f v_f^2 T \frac{1}{2\mu\Delta_0}. \qquad (50)$$

For heat flow along the $c$-axis the results for these same pairing states differ significantly. First consider the hybrid-I state. The angular average now reduces to $\frac{1}{2}\cdot\int_0^\pi d\vartheta \sin\vartheta \cos^2\vartheta\, \gamma^2(\Delta^2+\gamma^2)^{-3/2}$. The gap opens up linearly at the positions of the point nodes at the poles. In the limit $\gamma\to 0$, the integrand diverges as $|\vartheta-\pi/2|^{-1}$ near the line node, and as $\vartheta^{-2}$ near the point nodes. Thus the integral is dominated by the contribution from the point nodes (note, however, that $\gamma$ is mainly determined by the line node). Linearizing the gap near the point node $\Delta(\vartheta)\simeq\mu_1\Delta_0\vartheta$, we obtain for the low-temperature limit of the $c$-axis component of the thermal conductivity

$$\kappa_\perp(T) \simeq \frac{\pi^2}{3}N_f v_f^2 T \frac{\gamma}{\mu_1^2\Delta_0^2}, \qquad (51)$$

which is nonuniversal, and generally much less than $\kappa_\parallel$ by a factor of order $(\gamma/\Delta_0)$. To differentiate gaps with point nodes of different orders we classify them by their first nonvanishing derivative at the nodal points [i.e., for an $n$-th order point node, $\Delta\simeq\mu_n\Delta_0\vartheta^n$].

In the case of the hybrid-II gap the $c$-axis transport is more subtle. The gap opens quadratically with angle near the point nodes at the poles, i.e., $\Delta(\vartheta)\simeq\mu_2\Delta_0\vartheta^2$, and once again we obtain a universal result for the zero-temperature thermal conductivity,

$$\kappa_\perp(T) \simeq \frac{\pi^2}{3}N_f v_f^2 T \frac{1}{2\mu_2\Delta_0}. \qquad (52)$$

The ground state for the $E_{2u}$ model[39] of UPt$_3$ is an example of a hybrid-II state. Thus, an important feature of this model is that both components of the thermal conductivity tensor have universal values in the limit $T\to 0$. If we use the polynomial form for the $E_{2u}$ order parameter, $\Delta(\vec{p}_f)\sim p_{fz}(p_{fx}+ip_{fy})^2$, then the slope of the gap at the line node and the curvature of the gap at the point node are identical, $\mu\equiv\mu_2=3\sqrt{3}/2$. Thus, for a spherical Fermi surface we find that all the eigenvalues of $\overleftrightarrow{\kappa}$ are identical in the limit $T\to 0$. This result is consistent with the result reported in Ref. 47. However, the isotropy of $\kappa_{ij}$ is a peculiarity of the polynomial basis functions for the $E_{2u}$ representation, and of course the spherical Fermi surface.

Finally, we consider a gap with cubic point nodes at the poles, i.e., $\Delta(\vartheta)\simeq\mu_3\Delta_0\vartheta^3$. An example is the hybrid-III-A state with $E_{1g}$ symmetry, $\Delta(\vec{p}_f) = 16/(3\sqrt{3})\,\Delta_0\, p_{fz}(p_{fx}^2+p_{fy}^2)(p_{fx}+ip_{fy})$, which has a linear line node in the basal plane, but also cubic point nodes at $\vartheta=0,\pi$. The $B_{1g}$ and $B_{2g}$ states of a hexagonal crystal possess cubic point nodes along the $c$-axis if one assumes an analytic expansion of $\Delta$ in terms of $\vec{p}_f$. The odd-parity $B_{1u}$ and $B_{2u}$ states also possess cubic point nodes if we restrict the spin quantization axis to $\hat{d}=\hat{c}$; however, more general spin states do not possess cubic point nodes. It is also worth noting that cubic point nodes are expected for a large number of superpositions of two 1D representations, as in the "accidental degeneracy" models[51] for UPt$_3$ [cf. Refs. 52 and 39]. There is an important difference between the 2D hybrid-III-A state and the various 1D representations of the hexagonal group. The cubic point nodes of the 1D or mixed symmetry ground states are generally combined with line nodes connecting the point nodes at opposite poles [An exception is the degenerate $B_{2g}+iB_{1g}$ (hybrid-III-B) state]. Thus, there is a higher density of excitations for the 1D states than for the hybrid-III-A state in the vicinity of the cubic point nodes. The large density of gapless excitations in the vicinity of a cubic point node leads to a relatively large nonuniversal value for the low-temperature limit of the thermal conductivity in the $c$ direction. For the case of a pure cubic point node we obtain

$$\kappa_\perp(T) \simeq \frac{\pi^2}{3} N_f v_f^2 T \frac{2\pi\sqrt{3\pi}}{27\mu_3 \Delta_0 \Gamma(\frac{2}{3})\Gamma(\frac{5}{6})} \left(\frac{\mu_3 \Delta_0}{\gamma}\right)^{\frac{1}{3}} \quad (53)$$

in the limit $T \to 0$. Note, that the slope of the $c$-axis thermal conductivity is enhanced relative to the universal limit of the quadratic point node by $(\mu_3 \Delta_0/\gamma)^{1/3}$. Of course this value is limited at large $\gamma$ by the normal state value.

Although we have so far assumed an isotropic or cylindrical Fermi surface, many of the results are more general, or simply extended to include uniaxial anisotropy. Also, one can generalize the results to an arbitrary anisotropic Fermi surface with anisotropic impurity scattering of the form $u_{\vec{p}_f \vec{p}_f'} = u_0\, \eta(\vec{p}_f)\eta(\vec{p}_f')$, where $\eta(\vec{p}_f)$ is any basis function with the full symmetry of the Fermi surface. In addition, for the zero temperature limit only the density of states and Fermi velocities at the appropriate nodes are involved. Thus, the results for the zero temperature slopes of the thermal conductivity are easily extended to include Fermi surface anisotropy by replacing $N_f v_f^2$ by the value of this quantity at the position of the node, or the relevant one-dimensional average in the case of a line node.

### B. Low Temperature Corrections

Impurity scattering in an unconventional superconductor with line nodes leads to a finite density of zero-energy excitations. The bandwidth of these impurity bound states is of order $\gamma$. The leading order finite temperature corrections to the transport coefficients are obtained by a Sommerfeld expansion of Eqs. (30) and (31) for the thermal and electrical conductivities. The key point is that the impurity-renormalized excitation spectrum has according to the symmetry relation (32) a low-energy expansion of the form

$$\tilde{\epsilon}^{R,A}(\epsilon) \approx \pm i(\gamma + b\epsilon^2) + a\epsilon, \quad (54)$$

with real coefficients $a, b$ and $0 < \gamma \ll \Delta_0$. Expanding the integrands in Eqs. (30) and (31) to $\mathcal{O}[\epsilon^2]$ gives the Sommerfeld expansion for the components of $\kappa_{ij}$ and $\sigma_{ij}$,

$$\kappa_{ii}(T) \simeq \frac{N_f v_f^2}{4T^2} \int d\epsilon\, \epsilon^2 \operatorname{sech}^2\left(\frac{\epsilon}{2T}\right) \Big\{\gamma^2 I_{3/2}$$
$$+ \epsilon^2 \Big[2\gamma b\, I_{3/2} + \left(\frac{5}{2}a^2\gamma^2 - 3b\gamma^3\right)I_{5/2}$$
$$- \frac{5}{2}a^2\gamma^4 I_{7/2}\Big]\Big\}, \quad (55)$$

$$\operatorname{Re}\sigma_{ii}(T, \omega\to 0) \simeq \frac{e^2 N_f v_f^2}{4T} \int d\epsilon\, \operatorname{sech}^2\left(\frac{\epsilon}{2T}\right) \Big\{\gamma^2 I_{3/2}$$
$$+ \epsilon^2 \Big[2\gamma b\, I_{3/2} + \left(\frac{15}{2}a^2\gamma^2 - 3b\gamma^3\right)I_{5/2}$$
$$- \frac{15}{2}a^2\gamma^4 I_{7/2}\Big]\Big\}, \quad (56)$$

where $I_\nu = \left\langle \hat{v}_{fx}^2 (\Delta^2 + \gamma^2)^{-\nu} \right\rangle$ for the in-plane components and $I_\nu = \left\langle \hat{v}_{fz}^2 (\Delta^2 + \gamma^2)^{-\nu} \right\rangle$ for the $c$-axis components. The different numerical coefficients for the terms involving $a^2$ are due to the difference in the coherence factors for electrical and thermal transport. For $\nu = 3/2$ these integrals have been evaluated: $I_{3/2} = 2/(\pi\Delta_0 \gamma^2)$ for the $d_{x^2-y^2}$ state and $I_{3/2} = 1/(2\Delta_0 \gamma^2)$ for the states with a line of nodes at $\vartheta = \pi/2$. We find in all cases $I_{5/2}/I_{3/2} = 2\gamma^2/3$ and $I_{7/2}/I_{3/2} = 8\gamma^4/15$. It is remarkable that $b$ always drops out. After performing the $\epsilon$ integrals — $\int d\epsilon\, \epsilon^{2n} \operatorname{sech}^2(\epsilon/2T) = b_n \pi^{2n} T^{2n+1}$, with $b_0 = 4, b_1 = 4/3, b_2 = 28/15$ — we obtain for the $d_{x^2-y^2}$ state

$$\kappa_\parallel(T) \simeq \frac{\pi^2}{3}T\, \frac{2N_f v_f^2}{\pi\mu\Delta_0}\left(1 + \frac{7\pi^2}{15}\frac{a^2 T^2}{\gamma^2}\right), \quad (57)$$

and

$$\operatorname{Re}\sigma_\parallel(T, \omega\to 0) \simeq e^2 \frac{2N_f v_f^2}{\pi\mu\Delta_0}\left(1 + \frac{\pi^2}{3}\frac{a^2 T^2}{\gamma^2}\right), \quad (58)$$

to leading order in $aT/\gamma$. For the other pairing states (polar, hybrid-I, hybrid-II) the in-plane transport coefficients are obtained from Eqs. (57)-(58) by multiplying by $\pi/4$. The coefficient $a$ is strongly dependent on the phase shift. For resonant scattering $a = 1/2$, independent of the specific pairing state. Note, that Eq. (58) agrees with the result of Hirschfeld et al. in the resonant limit.[49] In the Born limit $a = \pi\mu\Delta_0\tau_0/2$ for the $d_{x^2-y^2}$ state, and $a = 2\mu\Delta_0\tau_0$ for the other states (ignoring the special case of quadratic or cubic point nodes). Since we assume $\Delta_0\tau_0 \gg 1$, we always have for weak scattering $a \gg 1$.

Finally, note that the finite temperature correction to the Wiedemann-Franz ratio becomes

$$L(T) = \frac{\kappa_\parallel(T)}{T\operatorname{Re}\sigma_\parallel(T,\omega\to 0)} \simeq L_S\left(1 + \frac{2\pi^2}{15}\frac{a^2 T^2}{\gamma^2}\right), \quad (59)$$

which increases with temperature for $T \ll T^* \sim \gamma$. This behavior arises from two sources: (i) the density of states, which is finite at $\epsilon = 0$ with $N(0) \sim N_f\,(\gamma/\Delta_0)$, depends strongly on energy for $\epsilon \gtrsim \gamma$, and (ii) the difference in the coherence factors for thermal and electrical conduction, which also depend on $\epsilon$. Note that if scattering is weak, or if the material is very clean, then the very low temperature regime may be difficult to achieve in practice.

### V. NUMERICAL RESULTS

More detailed information can be obtained from numerical evaluations of the transport coefficients over the full temperature range below $T_c$. The numerical results reported here were obtained by computing the equilibrium propagator, self-energy, and order parameter self-consistently for the four pairing models — $d_{x^2-y^2}$, polar, hybrid-I and hybrid-II — then using these results as input to numerically evaluate Eqs. (30)-(31) for the transport coefficients. We assumed a spherical Fermi surface except for the 2D $d_{x^2-y^2}$ state for which we used a cylindrical Fermi surface.

Our numerical results agree with those of previous authors[24,46,49,42,53–56] in those cases where a direct comparison is possible.

Figure 1 shows the results of the in-plane thermal conductivity for the four pairing states in the resonant scattering limit, i.e., for a normalized scattering cross section of $\bar{\sigma} = \sin^2 \delta_0 \equiv 1$. All curves exhibit the qualitative behavior of a superconductor with lines of nodes as discussed by many authors. It is remarkable that the curves for the 2D $d_{x^2-y^2}$ state and the 3D $E_{2u}$-state are essentially identical.

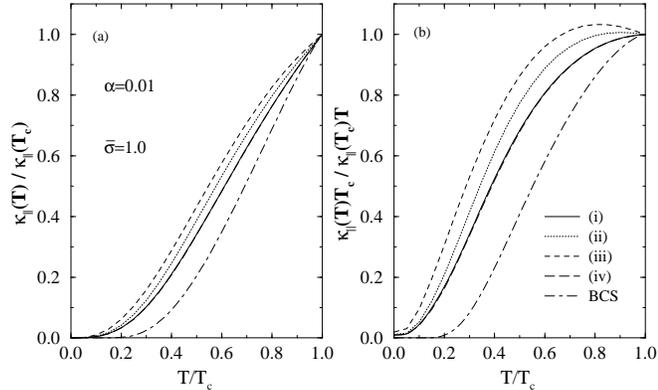

FIG. 1. Thermal conductivity $\kappa_\parallel$ vs. temperature for unconventional superconductors in the unitarity limit ($\bar{\sigma} = 1$) with a dimensionless scattering rate $\alpha = 0.01$. The different pairing states are: (i) $d_{x^2-y^2}$, (ii) polar, (iii) hybrid-I, and (iv) hybrid-II as described in the text. For comparison the result for an isotropic BCS superconductor is shown. Panel (a) displays the in-plane or basal plane $\kappa_\parallel$ normalized to its value at $T_c$. Panel (b) displays the ratio $\kappa_\parallel/T$ normalized to its value at $T_c$. Note the finite intercept of $\kappa_\parallel/T$ for the unconventional pairing states.

In Fig. 2 we plot for a $d_{x^2-y^2}$ pairing state $[\kappa_\parallel(T)T_c/\kappa_\parallel(T_c)T]$ vs. $T/T_c$ for several (normalized) scattering cross sections $\bar{\sigma} = \sin^2 \delta_0$ and (normalized) scattering rates $\alpha = 1/(2\pi T_{c0}\tau_0)$. A consequence of the universal limit for $T \to 0$ is that the ratio

$$\lim_{T \to 0} \frac{\kappa_\parallel(T) T_c}{\kappa_\parallel(T_c) T} \simeq \frac{1}{\pi \tau_0 \Delta_0(0)} = \alpha \frac{2 T_{c0}}{\Delta_0(0)} \qquad (60)$$

scales linearly with the scattering rate parameter $\alpha$, provided $\alpha$ is significantly less than the critical pair-breaking value ($\alpha_{cr} \approx 0.28$), and is independent of the scattering strength, $\bar{\sigma}$. At $T \gtrsim T^*$ and for weak scattering Arfi et al.[54] have shown that the ratio $[\kappa_\parallel(T)T_c]/[\kappa_\parallel(T_c)T] \propto (1-\bar{\sigma})$ strongly depends on the scattering phase shift $\delta_0$. This explains (i) the sudden drop of $[\kappa_\parallel(T) T_c]/[\kappa_\parallel(T_c)] T$ in Fig. 2(a) at ultra-low temperatures for weak scattering, where the universal limit is achieved only for $T \lesssim T^* \sim \Delta_0 \exp(-1/\alpha)$, and (ii) the scaling of the zero temperature intercept in Fig. 2(b). To address the various power law behaviors of $\kappa_\parallel(T)$ in different temperature regions and for different scattering rates and scattering cross sections, we show $\kappa_\parallel(T)/\kappa_\parallel(T_c)$ in Figs. 2(c,d) in a log-log plot for the same parameters as in Figs. 2(a,b), respectively. The temperature dependence of the electronic thermal conductivity obeys a $T^3$ variation above a critical temperature $T^* \sim \gamma$ in clean superconductors and in the strong scattering regime. Below $T^*$ it approaches the limiting $T$ behavior. Weak scattering leads to an approximately linear temperature dependence over a large portion of the temperature range. However, the ratio $\kappa_\parallel/T$ changes drastically in clean superconductors below the exponentially small crossover temperature $T^*$, where it approaches its linear low-temperature asymptote.

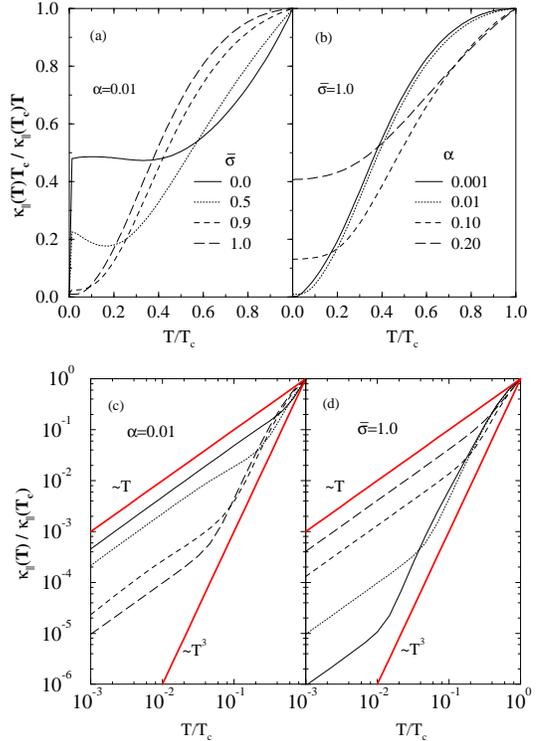

FIG. 2. Thermal conductivity of a 2D $d_{x^2-y^2}$-wave superconductor. The temperature dependence of $\kappa_\parallel(T)T_c/\kappa_\parallel(T_c)T$ vs. $T/T_c$ is displayed in (a) for different scattering cross sections, $\bar{\sigma} = \sin^2 \delta_0$, and for a fixed scattering rate $\alpha = 0.01$. In panel (b): the temperature dependence is plotted for different scattering rates $\alpha$ in the unitarity limit ($\bar{\sigma} = 1.0$). Panels (c) and (d) are the same calculations as panels (a) and (b), respectively, plotted on a log-log scale in order to exhibit the low temperature power laws and crossovers. The thick dotted lines are guidelines to $T$ and $T^3$ power laws.

As $T \to 0$, we indeed find the universal behavior as discussed in the previous section. To show the approach to the universal limits at low temperatures, we computed the electrical and thermal conductivity at low and ultralow temperatures for an intermediate scattering rate $\alpha = 0.1$, chosen because the temperature range of universality is exponentially small for weak scattering $\propto \exp(-1/\alpha)$. The results are shown in Fig. 3 where the electrical and thermal conductivities have been normalized by their corresponding universal limits, $\sigma_0 \equiv \sigma_\parallel(T \to 0, \omega \to 0)$ and $\kappa_0(T) \equiv T \frac{d\kappa_\parallel}{dT}(0)$. The corresponding Lorenz ratio is shown in Fig. 3(c). Notice the logarithmic scale in temperature. In agreement with the analytical results, numerical calculations show that Re $\sigma_\parallel(T, \omega \to 0)$, $\kappa_\parallel(T)/T$ and $L(T)$ *increase* with temperature near $T = 0$. The crossover temperature to the universal regime is exponentially small for small phase shifts, and the Lorenz number quickly drops for these smaller phase shifts after the initial rise.

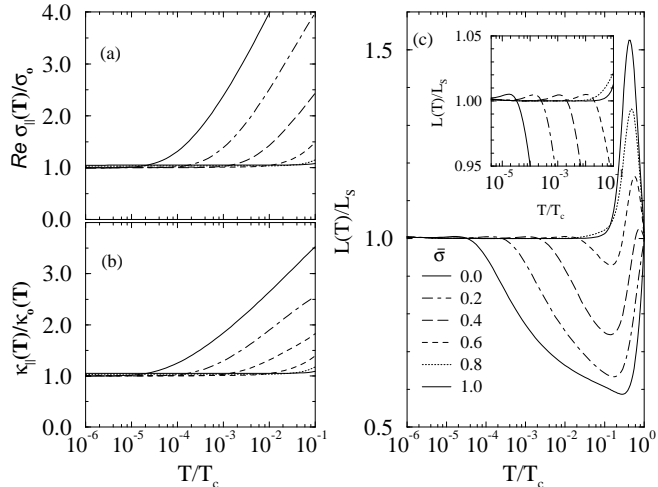

FIG. 3. Transport coefficients $\text{Re}\,\sigma_\parallel(T, \omega \to 0)$ and $\kappa_\parallel(T)$ vs. $T$ of a 2D $d_{x^2-y^2}$-wave superconductor for different scattering cross sections ($\bar{\sigma} = 0 \ldots 1$) and a fixed intermediate (normalized) scattering rate $\alpha = 0.1$. (a) The normalized electrical conductivity $\text{Re}\,\sigma_\parallel(T, \omega)/\sigma_0$ at a very small frequency $\omega = 10^{-6}\Delta_0(0)$. (b) The normalized thermal conductivity $\kappa_\parallel(T)/\kappa_0(T)$. (c) The normalized Lorenz ratio $L(T)/L_S$. Note the significant difference in $L(T)/L_S$ for unitarity vs. Born scattering. Inset: Blow-up of the *universal* behavior at ultra-low temperatures showing the approach to $L_S$ at $T \to 0$.

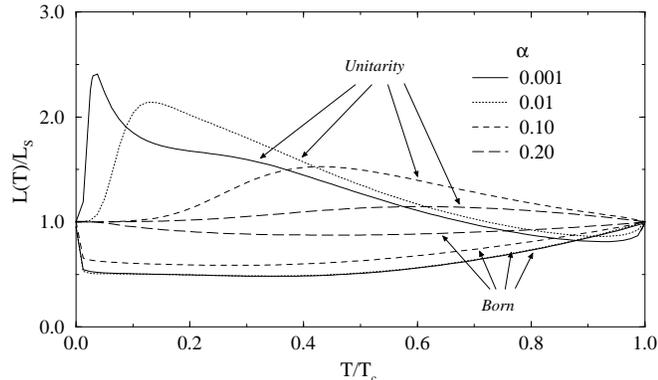

FIG. 4. The normalized Lorenz ratio $L(T)/L_S$ of a 2D $d_{x^2-y^2}$-wave superconductor for different (normalized) scattering rates $\alpha$ in the Born ($\bar{\sigma} = 0$) and unitarity ($\bar{\sigma} = 1$) limit. $L(T)/L_S$ clearly separates between scattering in the weak limit (lower four curves) and the strong limit (upper four curves).

In Fig. 4 we show the Lorenz ratio, $L(T)/L_S$, over the full temperature range for scattering in both the weak (Born) limit and the resonant (unitarity) limit. The deviations from the Sommerfeld value, $L_S$, are clearly separated between the Born and unitarity limits. This effect is most pronounced in nearly pure systems. At temperatures close to $T_c$, and in the clean limit with strong scattering, the Lorenz ratio is slightly reduced due to a small coherence peak in the electrical conductivity. In the Born limit $L(T)/L_S$ is always less than the normal state limit, which is opposite to that for unitarity scattering. At sufficiently low temperatures all curves converge to the same *universal* limit.

In superconductors with a large concentration of (resonant) scatterers the Wiedemann-Franz law is fulfilled throughout the entire temperature range, a result which is obvious from the fact that nonmagnetic impurities lead to pair-breaking in unconventional superconductors, and as the impurity lifetime $\tau_0$ approaches the critical pair-breaking value, the transport properties approach those of the normal metal.

## VI. COMPARISON WITH EXPERIMENTS

### A. High $T_c$ cuprates

Several measurements of the thermal[57,58] and electrical[59,60] conductivity on single crystals of high $T_c$ cuprates have been reported. In-plane thermal conductivity measurements show the presence of a 'low-temperature' $T^3$ term, which has been interpreted as boundary scattering of phonons on crystal faces, as well as a linear term at very low temperatures, which has been attributed to uncondensed charge carriers [for a review see Ref. 1].

In Table II we list the material parameters for the cuprate superconductors which we used to estimate the slope of the thermal conductivity at zero temperature. In the analysis we used the Drude plasma frequency to determine $N_f v_f^2$, i.e., $\omega_p^2 = 4\pi e^2 N_f v_f^2$, the weak-coupling $d$-wave gap ratio, $\Delta_0(0) = 2.14\, k_B T_c$, and the universal lifetime $\tau_\Delta \simeq 1/(\pi \Delta_0)$ to rewrite the universal limit of the electrical conductivity as

$$\sigma_0 \simeq \frac{\omega_p^2 \tau_\Delta}{4\pi}. \quad (61)$$

This enables us to estimate the slope of the thermal conductivity from the WF law,

$$\lim_{T \to 0} (\kappa(T)/T) \simeq \sigma_0 L_S. \quad (62)$$

For Y-Ba-Cu-O$_{6.95}$ the conductivity at $T \to 0$ estimated from microwave experiments is $\sigma_a \approx 0.5\,(\mu\Omega\,\text{m})^{-1}$ along the a-axis and $\sigma_b \approx 0.7\,(\mu\Omega\,\text{m})^{-1}$ along the b-axis.[61] These values are close to the universal value obtained from Eq. (61) and Table II, $\sigma_0 \simeq 0.5 - 0.6\,(\mu\Omega\,\text{m})^{-1}$, and provide reasonable agreement between the slope of the thermal conductivity obtained from the Wiedemann-Franz ratio (62) and the experimental values listed in Table III.[62] However, the experimental coefficients vary by as much as a factor of $\sim 8$ among different samples of Y-Ba-Cu-O$_{6.95}$. The theoretical values for $\lim_{T \to 0} \kappa_\parallel / T$ lie in the range of experimental values except for La-Sr-Cu-O. Thus, if La-Sr-Cu-O is a $d$-wave superconductor resolution of this discrepancy would require weak-scattering. In which case it would be very difficult to determine the *universal* zero temperature slope of $\kappa_\parallel$, because the region where $\kappa_\|$ shows universal behavior is exponentially small. Thus, for Born scattering the low temperature extrapolation of $\kappa_\|/T$ will overestimate the universal limit.

TABLE II. Material parameters of optimally doped cuprate samples.

| Compound | $T_c$ (K) | $\omega_p$ (eV) | Refs. |
|---|---|---|---|
| La-Sr-Cu-O | 35 | 0.8 − 0.9 | 68, 69 |
| Y-Ba-Cu-O | 92 | 1.4 - 1.5 | 61, 70, 71 |
| Bi-Sr-Ca-Cu-O | 86 | 1.1 - 1.2 | 70, 72 |

TABLE III. Slopes of the low-temperature thermal conductivity for the cuprates. The theoretical value is the universal slope calculated for a $d_{x^2-y^2}$ pairing state in the limit of resonant scattering (see Sect. VI).

| $[\kappa/T]_{T\to 0}$ (mW/K$^2$m) | La-Sr-Cu-O | Y-Ba-Cu-O | Bi-Sr-Ca-Cu-O |
|---|---|---|---|
| theor. | 10 - 13 | 11 - 14 | 8 - 10 |
| exp. | 100 - 150 | 5 - 12, 40 | ∼ 6 |
| Refs. | 73, 74 | 1, 58 | 1 |

### B. The heavy fermion superconductor UPt$_3$

For UPt$_3$ and heat flow in the basal plane the thermal conductivity in the normal state obeys $\kappa_n(T)/T = 1/(a + bT^2)$, with $a = 0.25\,\mathrm{m\,K^2\,W^{-1}}$ and $b = 1.0\,\mathrm{m\,W^{-1}}$, and it is believed that the thermal conductivity is almost entirely electronic,[28] and that the $b$ term arises from electron-electron scattering. If we take the experimental data reported by Lussier, Ellman and Taillefer[28] for UPt$_3$ with $T_c \simeq 0.5\,\mathrm{K}$, an average Fermi velocity, $v_f = 5.5\,\mathrm{km/s}$, and a transport mean-free path of $l_0 = 220\,\mathrm{nm}$ [i.e., a transport time of $\tau_0 = 40\,\mathrm{ps}$], and combine it with the theoretical value for $\lim_{T\to 0} \kappa_\|(T)/T \approx 0.6\,\frac{T_c}{\mu\Delta_0(0)}\kappa_n(T_c)/T_c$, we can predict the universal low-temperature slope. For the E$_{2u}$ state we estimate a slope of $\lim_{T\to 0}\kappa_\|(T)/T \approx 0.1\,\kappa_n(T_c)/T_c \approx 2\,\mathrm{mW\,K^{-2}cm^{-1}}$, while the corresponding ratio for the E$_{1g}$ state is smaller by 20%. These estimated theoretical values of $\kappa_\|/T$ at $T=0$, are of the same magnitude as the experimental data at $T\approx 0.1\,T_c \simeq 50\,\mathrm{mK}$ with $\kappa_\|/T|_{exp} \simeq 2\,\mathrm{mW\,K^{-2}cm^{-1}}$.[29]

To distinguish these different pairing states it is necessary to probe a symmetry dependent quantity as the nonlinear Meißner effect,[63,50] or the $\frac{\pi}{3}$-phase shift Josephson effect.[39] Here we analyze the anisotropy ratio of the thermal conductivity. The experimentally observed anisotropy ratio

$$r_a \equiv \lim_{T\to 0} \frac{\kappa_\perp/\kappa_\|}{(\kappa_\perp/\kappa_\|)_n} \qquad (63)$$

approaches $r_a \simeq 0.43$ for $T \to 0$.[29] For an ellipsoidal Fermi surface the normal state anisotropy, e.g., $v_{f\perp} \neq v_{f\|}$, drops out of the anisotropy ratio in Eq. (63). Thus, we can use the formulas derived for an isotropic Fermi surface; the anisotropy is then determined by the anisotropy of the superconducting gap as defined by the gap anisotropy parameters, $\mu$, $\mu_1$, $\mu_2$, which measure the slope or curvature of the gap at a line or point node on the Fermi surface.[64]

The anisotropy ratios for the various order parameters are obtained from Table I. The E$_{1g}$ and E$_{2u}$ models have been discussed by many authors as models for the low temperature phases of UPt$_3$.[65,46,66,67,39,11] For the E$_{1g}$ (hybrid-I) state the anisotropy ratio is nonuniversal and given by

$$r_a \simeq \frac{2\gamma\mu}{\mu_1^2\Delta_0} \qquad (E_{1g})\,. \qquad (64)$$

For the standard E$_{1g}$ order parameter, $\Delta(\vec{p}) = 2\Delta_0 p_{fz}(p_{fx} + ip_{fy})$, we have $\mu = \mu_1 = 2$, and hence $r_a \simeq \gamma/\Delta_0$. Assuming the unitarity limit and $\Gamma_u = 0.1\,T_c$, we obtain $r_a \sim 0.2$. We can account for the experimental value of $r_a$ by adjusting $\mu$ and/or $\mu_1$. Although $\Gamma_u = 0.1\,T_c$ is consistent with Ref. 28 and other normal state measurements, it predicts a flattening of $\kappa_b$ for $T \lesssim 0.2\,T_c$, which is experimentally not observed. A smaller scattering rate, $\Gamma_u = 0.01\,T_c$, does a better job in accounting for the low temperature behavior of $\kappa_\|$, hence we obtain $r_a \sim 0.04$. Thus, it is not possible to account for the experimental anisotropy, $r_a \simeq 0.43$,[29] without a very large ratio $\mu/\mu_1^2$, which seems unphysical.

However, for the E$_{2u}$ (hybrid-II) state the anisotropy ratio becomes

$$r_a \simeq \mu/\mu_2 \qquad (E_{2u})\,, \qquad (65)$$

which is universal and independent of $\gamma$. Thus, we can easily fit the experimental value of $r_a$ for $T \to 0$ by choosing $\mu/\mu_2 \sim 0.43$, and then adjusting $\gamma$ to obtain the best fit to the low temperature behavior for $\kappa_\|(T)$.

Finally, we note that for the cubic point nodes the theoretical prediction for the normalized anisotropy ratio would be

$$r_a \simeq \frac{\mu}{\mu_3}(\mu_3\Delta_0/\gamma)^{1/3} \qquad (\mathrm{hybrid-III})\,. \qquad (66)$$

Therefore we generally expect $r_a > 1$, which clearly disagrees with experiment unless one chooses a very large $\mu_3$. Barring this unlikely scenario we can conclude that the anisotropy ratio rules out cubic (or higher order) point nodes for the gap in UPt$_3$.

### VII. CONCLUSION

In conclusion we have examined in detail the low-temperature behavior of the thermal and electrical conductivities and the Lorenz ratio for superconductors with line nodes. We considered several unconventional pairing states and found for some of the eigenvalues of the thermal conductivity tensor a *universal* value as $T\to 0$, similar to the electrical conductivity. Furthermore, we showed that the Wiedemann-Franz law is restored below a crossover temperature $T^*$. In clean systems we have $T^* \ll T_c$, while in 'dirty' superconductors the crossover temperature can be an appreciable fraction of $T_c$.

The estimates of the *universal* slope $\lim_{T\to 0} \kappa/T$, which we derived for the cuprates based on a $d_{x^2-y^2}$ order parameter, and for UPt$_3$, with a hybrid-I or -II gap, are comparable with experimentally reported values, except for the La-Sr-Cu-O materials. Assuming that La-Sr-Cu-O is an unconventional superconductor, the difference between theoretically and experimentally extrapolated values might be attributed to weak scatterers in this system. In UPt$_3$ the anisotropy ratio, $r_a = \lim_{T\to 0}(\kappa_\perp/\kappa_\parallel)/(\kappa_\perp/\kappa_\parallel)_n$ is consistent with an E$_{2u}$ gap and a universal value for $r_a$, or with an E$_{1g}$ gap and a nonuniversal value for $r_a$ determined by the impurity concentration. Further experiments on UPt$_3$ with controlled impurity concentrations should easily distinguish these two models

# ACKNOWLEDGMENTS

We would like to thank M. Palumbo for discussions and for valuable contributions to the numeric, and also Yu. Barash for important discussions on the thermal conductivity of unconventional superconductors. This research was partially supported by the National Science Foundation through the Northwestern University Materials Science Center, grant number DMR 91-20521, the Science and Technology Center for Superconductivity, grant number DMR 91-20000, and the Deutsche Forschungsgemeinschaft. D.R. and J.A.S. acknowledge joint support from the Alexander von Humboldt Stiftung and the DFG.

# APPENDIX: SOLUTIONS TO THE LINEARIZED QUASICLASSICAL TRANSPORT EQUATIONS

The deviations from local equilibrium, $\delta\hat{g}^X = \hat{g}^X - \hat{g}_0^X$, and $\delta\hat{\sigma}^X = \hat{\sigma}^X - \hat{\sigma}_0^X$ with $X \in \{R, A, K\}$, satisfy the linearized equations

$$\left[\delta\hat{g}^{R,A}, \hat{h}^{R,A}\right]_\circ = i\partial\hat{g}_0^{R,A} + \left[\hat{g}_0^{R,A}, \hat{\sigma}_{ext} + \delta\hat{\sigma}^{R,A}\right]_\circ, \tag{A1}$$

and

$$\hat{h}^R \circ \delta\hat{g}^K - \delta\hat{g}^K \circ \hat{h}^A - \hat{\sigma}_0^K \circ \delta\hat{g}^A + \delta\hat{g}^R \circ \hat{\sigma}_0^K =$$
$$-i\partial\hat{g}_0^K + \left(\hat{\sigma}_{ext} + \delta\hat{\sigma}^R\right) \circ \hat{g}_0^K + \delta\hat{\sigma}^K \circ \hat{g}_0^A$$
$$-\hat{g}_0^K \circ \left(\hat{\sigma}_{ext} + \delta\hat{\sigma}^A\right) - \hat{g}_0^R \circ \delta\hat{\sigma}^K, \tag{A2}$$

where $\hat{h}^{R,A} = \epsilon\hat{\tau}_3 - \hat{\sigma}_0^{R,A}$ and $\partial = \vec{v}_f \cdot \vec{\nabla}$. Note, that the equation for the Keldysh propagator, $\delta\hat{g}^K$, is coupled to the deviations of the retarded and advanced functions, $\delta\hat{g}^{R,A}$. We decouple these equations by using the equilibrium relations, Eqs. (23) and (24), and by introducing Eliashberg's anomalous propagator, $\delta\hat{g}^a$, and self-energy, $\delta\hat{\sigma}^a$, defined by

$$\delta\hat{g}^K = \delta\hat{g}^R \circ \Phi_0 - \Phi_0 \circ \delta\hat{g}^A + \delta\hat{g}^a, \tag{A3}$$
$$\delta\hat{\sigma}^K = \delta\hat{\sigma}^R \circ \Phi_0 - \Phi_0 \circ \delta\hat{\sigma}^A + \delta\hat{\sigma}^a. \tag{A4}$$

After eliminating $i\partial\hat{g}_0^R$ and $i\partial\hat{g}_0^A$ using Eq. (A1), the transport equation for $\delta\hat{g}^K$ is transformed into a transport equation for $\delta\hat{g}^a$,

$$\hat{h}^R \circ \delta\hat{g}^a - \delta\hat{g}^a \circ \hat{h}^A$$
$$+ \hat{g}_0^R \circ (i\partial\Phi_0) - (i\partial\Phi_0) \circ \hat{g}_0^A + \hat{g}_0^R \circ \delta\hat{\sigma}^a - \delta\hat{\sigma}^a \circ \hat{g}_0^A$$
$$+ [\hat{\sigma}_{ext}, \Phi_0]_\circ \circ \hat{g}_0^A + \hat{g}_0^R \circ [\Phi_0, \hat{\sigma}_{ext}]_\circ$$
$$= \left[(\hat{h}^R + \hat{\sigma}_0^R) \circ \Phi_0 - \Phi_0 \circ (\hat{h}^A + \hat{\sigma}_0^A)\right] \circ \delta\hat{g}^A$$
$$- \delta\hat{g}^R \circ \left[(\hat{h}^R + \hat{\sigma}_0^R) \circ \Phi_0 - \Phi_0 \circ (\hat{h}^A + \hat{\sigma}_0^A)\right]. \tag{A5}$$

The terms on the right hand side of Eq. (A5) vanish identically because

$$\hat{h}^A + \hat{\sigma}_0^A = \hat{h}^R + \hat{\sigma}_0^R = \epsilon\hat{\tau}_3. \tag{A6}$$

Thus, the equation for the anomalous propagator, $\delta\hat{g}^a$, becomes

$$\hat{h}^R \circ \delta\hat{g}^a - \delta\hat{g}^a \circ \hat{h}^A =$$
$$(i\partial\Phi_0) \circ \hat{g}_0^A - \hat{g}_0^R \circ (i\partial\Phi_0) + \delta\hat{\sigma}^a \circ \hat{g}_0^A - \hat{g}_0^R \circ \delta\hat{\sigma}^a$$
$$- [\hat{\sigma}_{ext}, \Phi_0]_\circ \circ \hat{g}_0^A - \hat{g}_0^R \circ [\Phi_0, \hat{\sigma}_{ext}]_\circ. \tag{A7}$$

The transport equations for $\delta\hat{g}^{R,A,a}$ are solved by noting that the local equilibrium propagators have the form

$$\hat{g}_0^{R,A} = \frac{\tilde{\epsilon}^{R,A}\hat{\tau}_3 - \hat{\tilde{\Delta}}^{R,A}}{C^{R,A}}, \tag{A8}$$

and that

$$\hat{h}^{R,A} = C^{R,A}\hat{g}_0^{R,A} + D^{R,A}\hat{1}, \tag{A9}$$

where

$$C^{R,A} = -\frac{1}{\pi}\sqrt{|\tilde{\Delta}^{R,A}|^2 - (\tilde{\epsilon}^{R,A})^2}, \tag{A10}$$

and the functions $\tilde{\epsilon}^{R,A}$, $\tilde{\Delta}^{R,A}$, and $D^{R,A}$ are defined in terms of the equilibrium self-energy

$$\hat{\sigma}_0^{R,A} = (\epsilon - \tilde{\epsilon}^{R,A})\hat{\tau}_3 + \hat{\tilde{\Delta}}^{R,A} + D^{R,A}\hat{1}. \tag{A11}$$

Finally, Eq. (A7) is solved by using the normalization condition for $\delta \hat{g}^a$,

$$\hat{g}_0^R \circ \delta \hat{g}^a + \delta \hat{g}^a \circ \hat{g}_0^A = 0, \tag{A12}$$

and the relations for $\hat{h}^{R,A}$ in Eq. (A9), to move $\delta \hat{g}^a$ to the right in the second term on the left side of Eq. (A7). Then (A7) turns into a form which can be solved by matrix inversion,

$$\delta \hat{g}^a = \left[ C_+^a \hat{g}_0^R + D_-^a \right]^{-1} \circ \Bigg( (i\partial \Phi_0) \circ \hat{g}_0^A - \hat{g}_0^R \circ (i\partial \Phi_0)$$
$$+ \delta \hat{\sigma}^a \circ \hat{g}_0^A - \hat{g}_0^R \circ \delta \hat{\sigma}^a$$
$$- [\hat{\sigma}_{ext}, \Phi_0]_\circ \circ \hat{g}_0^A - \hat{g}_0^R \circ [\Phi_0, \hat{\sigma}_{ext}]_\circ \Bigg), \tag{A13}$$

where

$$C_+^a(\vec{p}_f; \epsilon, \omega) = C^R(\vec{p}_f; \epsilon_+) + C^A(\vec{p}_f; \epsilon_-), \tag{A14}$$
$$D_-^a(\vec{p}_f; \epsilon, \omega) = D^R(\vec{p}_f; \epsilon_+) - D^A(\vec{p}_f; \epsilon_-) \tag{A15}$$

with $\epsilon_\pm = \epsilon \pm \omega/2$. Using the normalization condition (26) we write the inverse matrix as

$$\left[ C \hat{g}_0^{R,A} + D \right]^{-1} = -\frac{C \hat{g}_0^{R,A} - D}{\pi^2 C^2 + D^2}. \tag{A16}$$

Furthermore it is useful to introduce

$$C_+^{R,A}(\vec{p}_f; \epsilon, \omega) = C^{R,A}(\vec{p}_f; \epsilon_+) + C^{R,A}(\vec{p}_f; \epsilon_-), \tag{A17}$$
$$D_-^{R,A}(\vec{p}_f; \epsilon, \omega) = D^{R,A}(\vec{p}_f; \epsilon_+) - D^{R,A}(\vec{p}_f; \epsilon_-), \tag{A18}$$

and to obtain the solutions for $\delta \hat{g}^{R,A}$ by analogous steps,

$$\delta \hat{g}^{R,A} = \left[ C_+^{R,A} \hat{g}_0^{R,A} + D_-^{R,A} \right]^{-1} \circ \left( -i\partial \hat{g}_0^{R,A} + \left[ \hat{\sigma}_{ext} + \delta \hat{\sigma}^{R,A}, \hat{g}_0^{R,A} \right]_\circ \right). \tag{A19}$$

When combined with $\delta \hat{g}^a$ in Eq. (A13) according to (23) we obtain the general result for $\delta \hat{g}^K$,

$$\delta \hat{g}^K = \left[ C_+^R \hat{g}_0^R + D_-^R \right]^{-1} \circ \left( -i\partial \hat{g}_0^R + \left[ \hat{\sigma}_{ext} + \delta \hat{\sigma}^R, \hat{g}_0^R \right]_\circ \right) \circ \Phi_0$$
$$- \Phi_0 \circ \left[ C_+^A \hat{g}_0^A + D_-^A \right]^{-1} \circ \left( -i\partial \hat{g}_0^A + \left[ \hat{\sigma}_{ext} + \delta \hat{\sigma}^A, \hat{g}_0^A \right]_\circ \right)$$
$$+ \left[ C_+^a \hat{g}_0^R + D_-^a \right]^{-1} \circ \Bigg( (i\partial \Phi_0) \circ \hat{g}_0^A - \hat{g}_0^R \circ (i\partial \Phi_0) + \delta \hat{\sigma}^a \circ \hat{g}_0^A - \hat{g}_0^R \circ \delta \hat{\sigma}^a -$$
$$[\hat{\sigma}_{ext}, \Phi_0]_\circ \circ \hat{g}_0^A - \hat{g}_0^R \circ [\Phi_0, \hat{\sigma}_{ext}]_\circ \Bigg). \tag{A20}$$